\documentclass[%
 reprint,
groupedaddress,
nofootinbib,
 amsmath,amssymb,
 aps,
prc,
]{revtex4-2}

\usepackage{color} 
\usepackage{graphicx}
\usepackage{dcolumn}
\usepackage{bm}

\newcommand{\Ve}[1]{\ensuremath{\boldsymbol{#1}}}
\newcommand{\be}{\begin{equation}}
\newcommand{\ee}{\end{equation}}
\newcommand{\eq}[1]{\begin{align}#1\end{align}}
\usepackage{braket}
\usepackage[normalem]{ulem}
\usepackage{multirow}

\usepackage[normalem]{ulem}
\usepackage{comment}
\usepackage{xcolor}

\begin{document}


\title{Nonspherical Pauli-forbidden states in deformed halo nuclei:\\ Impact on the ${}^7\mathrm{Be}+p$ resonant states in the particle rotor model}

\author{Shin Watanabe}
\email[]{s-watanabe@gifu-nct.ac.jp}
\thanks{This work was conducted during a sabbatical leave at the Departamento de F\'isica At\'omica, Molecular y Nuclear, Universidad de Sevilla.}
\affiliation{National Institute of Technology (KOSEN), Gifu College, Motosu 501-0495, Japan}
\affiliation{RIKEN Nishina Center, Wako 351-0198, Japan}

\author{Antonio M. Moro}
\affiliation{Departamento de F\'isica At\'omica, Molecular y Nuclear, Universidad de Sevilla, Apartado 1065, E-41080 Sevilla, Spain}

\date{\today}

\begin{abstract}
\noindent{\bf Background:}
An important aspect of reducing nuclear many-body problems to few-body models is the presence of Pauli-forbidden (PF) states,
which are excluded in fully antisymmetrized calculations.
Insufficient treatments of PF states in deformed halo nuclei underscore the need for model refinement.

\noindent{\bf Purpose:}
We propose a new method utilizing Nilsson states as PF states in the orthogonality condition model,
and investigate the impact of PF states on the properties of resonant states.

\noindent{\bf Method:}
We investigate the scattering states of ${}^8\mathrm{B}$
within the particle rotor model (PRM) framework based on a deformed ${}^7\mathrm{Be}$ core and $p$ two-body model.
We compare several methods for eliminating PF states and test them with the experimental data.

\noindent{\bf Results:}
Our model successfully reproduces the experimental excitation function for elastic scattering
cross section by properly eliminating PF states.
The same calculation predicts the presence of a low-energy bump in the inelastic scattering
excitation function, although its position is overestimated by about 1 MeV compared to experimental data.

\noindent{\bf Conclusion:}
This study extends the applicability of the PRM, offering a comprehensive approach for exploring structures
and reactions of loosely bound nuclei such as ${}^8$B.
Future integration with the continuum discretized coupled channels method
promises to further advance the research.

\end{abstract}

\maketitle


\section{Introduction}

Beyond the conventional picture of spherical halos, recent experiments
have revealed the existence of deformed halo nuclei~\cite{Nak09,Nak11,Tak12,Kob14,Tak14}.
They are characterized not only by their extended neutron distributions
but also by the deformation of their core.
An accurate description of deformed halo nuclei is a frontier in modern nuclear physics,
integrating state-of-the-art models to elucidate the complex interplay between
a valence nucleon and a deformed core~\cite{Nav11,Min12,Wat14,Kas21,Kra23,Tak23}.
In the context of few-body models, the continuum discretized coupled channels (CDCC) method
has been significantly developed~\cite{Kam86,Aus87,Yah12,Hag22}.
Originally, CDCC was designed to describe reactions involving a deuteron and a target nucleus ($T$),
based on the $p+n+T$ three-body model.
Recently, CDCC has been extended to include core excitation effects,
leading to the extended version of CDCC (XCDCC)~\cite{Sum06,Sum06-2,Die14,Lay16,Die17}.
For example, XCDCC has been applied to ${}^{11}\mathrm{Be}$ scattering in the
${}^{10}\mathrm{Be}+n+T$ three-body model with ${}^{10}\mathrm{Be}$ core excitation,
revealing significant interference between the valence and core excitations in the breakup reactions~\cite{Cre11,Mor12,Mor12-2}.
Additionally, the inclusion of core excitation has been successfully achieved
 using the exact Faddeev formalism,
which has already been applied to all relevant reaction channels such as elastic, inelastic, transfer, and breakup~\cite{Del13,Del19}.
These efforts to accurately represent many-body systems through few-body models are a foundational task,
enabling us to interpret complicated phenomena from a simple perspective.

When reducing many-body problems to few-body models,
we often encounter the challenge of handling Pauli-forbidden (PF) states, especially for deformed nuclei.
The PF states are automatically excluded from the many-body wave function through antisymmetrization,
yet they require deliberate exclusion in few-body approaches. 
The problem becomes more involved when considering deformation effects, such as in the Nilsson model~\cite{Nil55,Rag95} or in the particle rotor model (PRM)~\cite{BM,Ura11,Ura12}.
Consider ${}^8$B in a ${}^7\mathrm{Be}+p$ two-body model as an example.
In the spherical shell model,
the last proton occupies the $0p_{3/2}$ orbital, which can accommodate up to four protons.
Consequently, there is no need for special treatment to eliminate PF states as long as we consider $p$-wave states.
However, when considering deformation, 
the $0p_{3/2}$ orbital splits into the [110 1/2] and [101 3/2] Nilsson
orbitals,\footnote{We consider the Nilsson orbital labeled by [$N n_3\Lambda$ $\Omega$] throughout this paper,
where $N$ is the principal quantum number denoting the major shell,
$n_3$ is the number of nodes along the $z'$ axis (symmetry axis),
and $\Lambda$ ($\Omega$) is the projection of the angular momentum
(total angular momentum) onto the $z'$ axis.}
which can accommodate up to two protons, respectively.
Therefore, the [110 1/2] orbital becomes forbidden, as it is already filled by two protons in the deformed ${}^{7}$Be core,
 necessitating careful treatment of PF states.

In this study, we focus on the refinement of the projectile wave function
of deformed halo nuclei using the PRM for further applications within the XCDCC reaction framework.
We demonstrate various treatments for excluding the PF states
and investigate their impact on the resonant-state properties.
Previous research presented a simple method to eliminate PF states in the deformed halo nucleus ${}^{31}\mathrm{Ne}={}^{30}\mathrm{Ne}+n$~\cite{Ura11,Ura12}.
In these works, the PF states are obtained as bound states,
 and eliminated after solving the Schr\"{o}dinger equation.
Due to its simplicity, this technique has been widely used in many studies~\cite{Mor12,Del16,Wat21,Pun23}.
However, as will be shown later, we find that this method cannot be directly applied
to ${}^{8}\mathrm{B}={}^{7}\mathrm{Be}+p$,
where the PF state emerges as a resonant state, complicating its exclusion.
To overcome this limitation, we propose using the Nilsson model to represent PF states
in the orthogonality condition model (OCM)~\cite{Sai68,Sai69,Hor77}. This method enables us to exclude
the PF states before solving the Schr\"{o}dinger equation.
This approach not only overcomes the limitations encountered in ${}^8$B calculation
but also offers the way for broader applications.
Integrating the projectile wave functions calculated from this method
with commonly used reaction frameworks, such as the XCDCC method (for breakup), and the distorted-wave Born approximation (DWBA) and the adiabatic distorted-wave approximation (ADWA)~\cite{Joh70} (for transfer reactions), could significantly improve our understanding and predictive capabilities in nuclear physics.

This paper is organized as follows.
In Sec.~\ref{sec:Formulation}, we first explain the PRM framework without considering the PF states.
This section is divided into three subsections, each explaining a different method for treating PF states.
Section~\ref{sec:PRM} provides a brief review of the treatment outlined in Refs.~\cite{Ura11,Ura12}.
In Sec.~\ref{sec:sph-PF}, we show the standard OCM using spherical PF states for comparison.
In Sec.~\ref{sec:def-PF}, we propose a new method using deformed PF states in the OCM.
Section~\ref{sec:results} presents the results obtained from these different models.
Finally, the paper concludes in Sec.~\ref{sec:summary}.

\section{Formulation} \label{sec:Formulation}
We consider the scattering states of ${}^{8}\mathrm{B}={}^{7}\mathrm{Be}+p$
using the two-body Hamiltonian with core excitation defined as
\be
H_1=T_{\Ve r}+V({\Ve r},{\Ve \xi})+h_\mathrm{core}({\Ve \xi}),\label{eq:H1}
\ee
where ${\Ve r}$ denotes the coordinate between proton and the core,
and ${\Ve \xi}$ is the internal coordinate of the core, i.e., direction of the symmetry axis of the deformed core.
The operator $T_{\Ve r}$ represents the kinetic energy and 
$h_\mathrm{core}$ is the internal Hamiltonian, governing the degree of freedom of the core:
$(h_\mathrm{core}-\epsilon_I)\phi_I=0$, where $\phi_I$ is the rotational wave function with the core spin $I$,
and $\epsilon_I$ is the corresponding eigenenergy.
In this study, an effective potential $V$ is assumed to be
\be
V({\Ve r},{\Ve \xi})=V_\mathrm{def}({\Ve r},{\Ve \xi})+V_{\ell s}(r){\Ve \ell}\cdot{\Ve s}+V_\mathrm{Coul}(r),
\ee
where $V_\mathrm{def}$ is a deformed potential, $V_{\ell s}$ is
the spin-orbit interaction with the angular momentum $\ell$ and proton spin $s=1/2$,
and $V_\mathrm{Coul}$ is the Coulomb interaction.
For simplicity, both $V_{\ell s}$ and $V_\mathrm{Coul}$ are treated as spherical potentials.

The total wave function of the ${}^7\mathrm{Be}+p$ system, $\Psi_{JM}$, is expanded as
\be
\Psi_{JM}({\Ve r},{\Ve \xi})=\sum_c \frac{u_c(r)}{r}\Phi_{c,JM}(\hat{\Ve r},{\Ve \xi})\label{eq:tot-wf}
\ee
with
\be
\Phi_{c,JM}(\hat{\Ve r},{\Ve \xi})=[\mathcal{Y}_{\ell j}(\hat{\Ve r})\otimes \phi_I({\Ve \xi})]_{JM},
\ee
where $J$ and $M$ denote the total angular momentum of the system and its $z$ component, respectively,
$j$ is the total angular momentum combining $\ell$ and $s$.
The function $\Phi_{c,JM}$ is the direct product of the spin-angular function $\mathcal{Y}_{\ell j m_j}$ 
and the core wave function $\phi_{I m_I}$, where $m_j$ ($m_I$) is the $z$ component of $j$ ($I$),
and $c$ indicates the channel $c=\{\ell j I\}$.
The function $u_c$ describes the relative motion between the proton and the core in the $c$ channel, and the set of $u_c$ is obtained by solving a set of coupled-channel equations
\be
[T_{r\ell}-\varepsilon_I]u_c(r)+\sum_{c'}V_{cc'}(r)u_{c'}(r)=0\label{eq:cc-eq}
\ee
under the appropriate boundary conditions.
Here, $T_{r\ell}$ denotes the kinetic energy,
$V_{cc'}(r)=\Braket{\Phi_{c,JM}|V|\Phi_{c',JM}}$ is the coupling potential,
and $\varepsilon_I$ is the relative energy
defined as $\varepsilon_I=\varepsilon-\epsilon_I$, with $\varepsilon$ being the total energy.

At this stage, the presence of PF states is not taken into account.

\subsection{Model I: Standard PRM} 
\label{sec:PRM}
\begin{figure}[htbp]
\includegraphics[width=0.9\linewidth,clip]{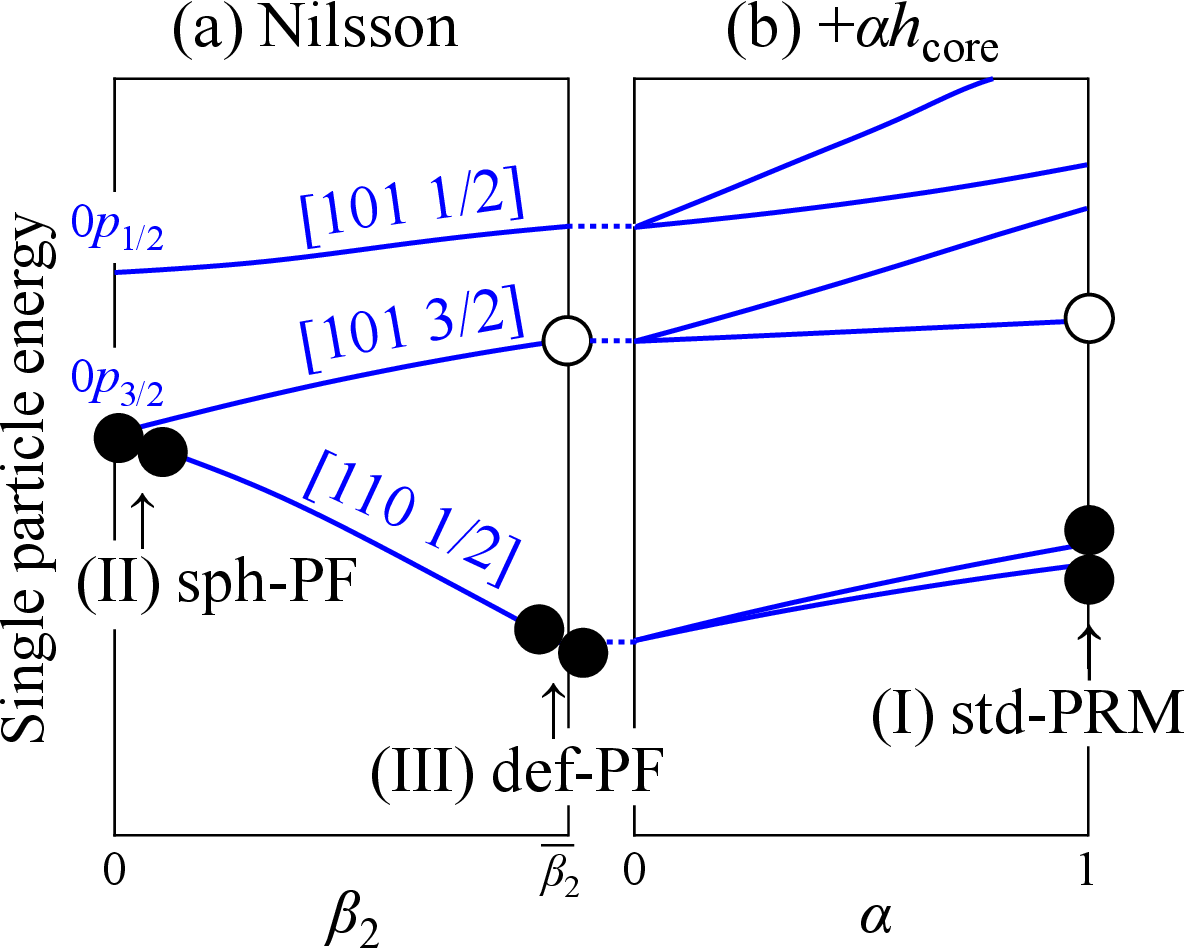}
\caption{Schematic illustration of the evolution of the single particle energy
by incorporating (a) deformation and (b) core excitation.
(a) corresponds to the Nilsson diagram, where the 
deformation parameter $\beta_2$ increases from 0 to a specific value $\bar{\beta}_2$.
(b) demonstrates the energy splitting by adding $\alpha h_\mathrm{core}$ to $H_\mathrm{Nil}$,
where $\alpha=0$ corresponds to the Nilsson model and $\alpha=1$ represents the PRM.
The filled circles denote the PF states while the open circle indicates the last proton.
Arrows labeled (I) std-PRM (standard PRM), (II) sph-PF (spherical PF states), and (III)
 def-PF (deformed PF states) are shown to explain various treatments of PF states. See text in detail. 
}
\label{fig:Nilsson_schematic}
\end{figure}

A simple method to eliminate the forbidden states in deformed halo nuclei is outlined in Refs.~\cite{Ura11,Ura12}.
We refer to this method as the ``standard PRM (std-PRM)" to distinguish it from the models described later.
First, we introduce the Nilsson Hamiltonian by omitting $h_\mathrm{core}$ from Eq.~\eqref{eq:H1} as
\be
H_\mathrm{Nil}=T_{\Ve r}+V({\Ve r},{\Ve \xi}). \label{eq:Nilsson}
\ee
By solving the Schr\"{o}dinger equation with $H_\mathrm{Nil}$, using the same form of wave functions [Eq.~\eqref{eq:tot-wf}],
we can conduct Nilsson model calculations within the PRM framework.
In this calculation, several states with different $J$ values, but having the same value of $\Omega$, become energetically degenerate if the model space \{$\ell j I$\} is sufficiently large.
Here, $\Omega$ denotes the projection of the total angular momentum onto the $z'$ axis in the body-fixed frame.
The anticipated behavior is schematically illustrated in Fig.~\ref{fig:Nilsson_schematic}(a),
in which the deformation parameter $\beta_2$ varies from 0 to a specific value $\bar{\beta}_2$.
Note that each Nilsson orbital is doubly degenerate due to time-reversal symmetry.
Following the standard procedure, we allocate two protons to each Nilsson orbital
starting from the lowest-energy level,
enabling us to identify the Nilsson orbital that accommodates the last unpaired proton.
Consequently, the [000 1/2] and [110 1/2] Nilsson orbitals are considered as the PF states.
In Fig.~\ref{fig:Nilsson_schematic}, PF states are represented by the filled circles,
while the last proton is indicated by the open circle for further explanation.

Subsequently, by gradually increasing the value of core excitation energies to the physical ones,
we can track the evolution of the Nilsson states.
For convenience, we introduce $\alpha h_\mathrm{core}$ to $H_\mathrm{Nil}$,
where $\alpha=0$ corresponds to the Nilsson model and $\alpha=1$ denotes the PRM.
This evolution is shown in Fig.~\ref{fig:Nilsson_schematic}(b).
At this stage, $\Omega$ is no longer a good quantum number, and each Nilsson orbital splits into two states.
Through this process, we can identify the states originating from the [000 1/2] and [110 1/2] Nilsson orbitals.
These states, indicated as ``(I) std-PRM" in Fig.~\ref{fig:Nilsson_schematic}(b), are interpreted as PF states, and subsequently excluded after solving the Schr\"{o}dinger equation with $H_1$.
The next-lowest-energy state, shown by the open circle in Fig.~\ref{fig:Nilsson_schematic}(b),
is then considered as the lowest-energy state in Model I.

\subsection{Model II: Spherical PF model} \label{sec:sph-PF}

Another way of eliminating PF states is to introduce
a Pauli blocking operator to Eq.~\eqref{eq:H1}, which is known as the orthogonality condition model (OCM)~\cite{Sai68,Sai69,Hor77}.
The Hamiltonian is defined as
\be
H_2=T+V+h_\mathrm{core}+V_\mathrm{PF}
\ee
with
\be
V_\mathrm{PF}=\sum_{i=1}^{N_\mathrm{PF}}\lambda_i\ket{\bar{\Psi}^{(i)}}\bra{\bar{\Psi}^{(i)}},\label{eq:VPF}
\ee
where the operator form is adopted for notational simplicity.
In Eq.~\eqref{eq:VPF}, $\bar{\Psi}^{(i)}$ represents the $i$th PF state and
 $N_\mathrm{PF}$ is the total number of PF states.
In practice, $\lambda_i=10^6$~MeV is chosen to ensure convergence.

In the standard OCM, spherical PF (sph-PF) states are  used
to define $\bar{\Psi}^{(i)}$ in Eq.~\eqref{eq:VPF}.
Therefore, in addition to our proposed model using the deformed PF (def-PF) states shown in the next subsection,
we also conduct calculations using the sph-PF states.
These sph-PF states (for given $J$ and $M$) can be explicitly defined using the same form as Eq.~\eqref{eq:tot-wf}:
\be
\bar{\Psi}_{JM}^{(i)}({\Ve r},{\Ve \xi})
=\sum_c \frac{g_c^{(i)}(r)}{r}\Phi_{c,JM}(\hat{\Ve r},{\Ve \xi}),\label{eq:PFstate}
\ee
where $g_c^{(i)}$ is the radial wave function.
Once $\bar{\Psi}_{JM}^{(i)}$ is determined, the set of coupled-channel equations for $u_c$ is derived as follows:
\eq{
&[T_{r\ell}-\varepsilon_I]u_c(r)+\sum_{c'}V_{cc'}(r)u_{c'}(r)\nonumber\\
&\hspace{5mm}+\sum_{c'}\int dr' W_{cc'}(r,r')u_{c'}(r')=0,\label{eq:NL-cc}
}
where $W_{cc'}(r,r')$ is the non-local potential defined as
\be
W_{cc'}(r,r')=\sum_{i=1}^{N_\mathrm{PF}}\lambda_i g_c^{(i)}(r)g_{c'}^{(i)*}(r').
\ee
These non-local coupled-channel equations are solved under the same boundary condition as for Eq.~\eqref{eq:cc-eq}.
The $R$-matrix code available in Refs.~\cite{Des16,Des10} is utilized.

In the actual ${}^7\mathrm{Be}+p$ calculation of the sph-PF model, 
the sph-PF states, equivalent to two $0p_{3/2}$ states, are determined
in the Nilsson model by setting $\beta_2=0.0001$ instead of zero.
This minimal $\beta_2$ is necessary to avoid possible configuration
mixing of the four $0p_{3/2}$ states, which could occur at  $\beta_2=0$
due to the degeneracy of $\epsilon_I=0$, as described by $[\mathcal{Y}_{p 3/2}\otimes \phi_I]_{JM}$.
Introducing a small $\beta_2$ value effectively resolves this degeneracy
and aligns the states, allowing for a reasonable definition
of sph-PF states corresponding to two [110 1/2] Nilsson states in the spherical limit.
These well-defined states are then employed as the PF states in Eq.~\eqref{eq:VPF},
which are indicated as ``(II) sph-PF" in Fig.~\ref{fig:Nilsson_schematic}.

\subsection{Model III: Deformed PF model} 
\label{sec:def-PF}
As a novel approach, we propose using def-PF states in the OCM, which we call the def-PF model.
In this model, we assume that $\bar{\Psi}^{(i)}$ in Eq.~\eqref{eq:VPF} should be deformed
rather than spherical. A natural choice is to use the $i$th Nilsson state,
which is indicated by ``(III) def-PF" in Fig.~\ref{fig:Nilsson_schematic}.
As discussed in Sec.~\ref{sec:PRM}, the Nilsson state (for given $J$ and $M$) can be obtained
by solving the Schr\"{o}dinger equation with $H_\mathrm{Nil}$ within the PRM framework.
Therefore, the Nilsson states $\bar{\Psi}_{JM}^{(i)}({\Ve r},{\Ve \xi})\equiv\braket{JM;{\Ve r},{\Ve \xi}|\bar{\Psi}^{(i)}}$
are explicitly defined in Eq.~\eqref{eq:PFstate}.
Once $\bar{\Psi}_{JM}^{(i)}$ is determined, the set of coupled-channel equations~\eqref{eq:NL-cc}
is solved in the same way as described in Sec.~\ref{sec:sph-PF}.

\section{Results}\label{sec:results}
\subsection{Model setting and structure analysis}
In the actual calculation, we start with the potential parameters for the spherical model shown in Ref.~\cite{Spa21}.
For simplicity, the core-spin orbit interaction, i.e., $V_{L I}$ in Ref.~\cite{Spa21}, is neglected.
For the deformed potential $V_\mathrm{def}$, we adopt the deformed Woods-Saxon potential with radius $R_0=2.391$ fm and diffuseness $a_0=0.535$ fm. The depth of this deformed Woods-Saxon potential, denoted as $V_\mathrm{def}^0$,
 will be determined later.
The Woods-Saxon volume type is employed for $V_{\ell s}$, with a depth $V_{LS}^0=-7$ MeV,
and the same radius $R_0=2.391$ fm and diffuseness $a_0=0.535$ fm.
Regarding the core state, we consider the negative-parity states from
$I^\pi=1/2^-$ to $13/2^-$ to clearly define the PF states.
For the energies $\epsilon_I$ corresponding to these states, we use the experimental data for the four lowest-energy states
($\epsilon_{3/2}=0$ MeV, $\epsilon_{1/2}=0.43$ MeV, $\epsilon_{7/2}=4.57$~MeV, and $\epsilon_{5/2}=6.73$~MeV).
For the higher-energy states, we use extrapolated values using the formula~\cite{Mar15,Mar19}:
\be
\epsilon_I=\epsilon^0+A\left[I(I+1)+a(-)^{I+1/2}(I+1/2)\right],
\ee
where the constants ($\epsilon^0$, $A$, and $a$) are determined from
the three lowest experimental energies:
$\epsilon^0=-0.564$~MeV, $A=0.486$~MeV, and the decoupling factor $a=-1.29$.
The deformation parameter $\beta_2$ is set to 0.586.

\begin{figure*}[ht]
\centering
\begin{minipage}{0.33\linewidth}
\includegraphics[width=0.9\linewidth,clip]{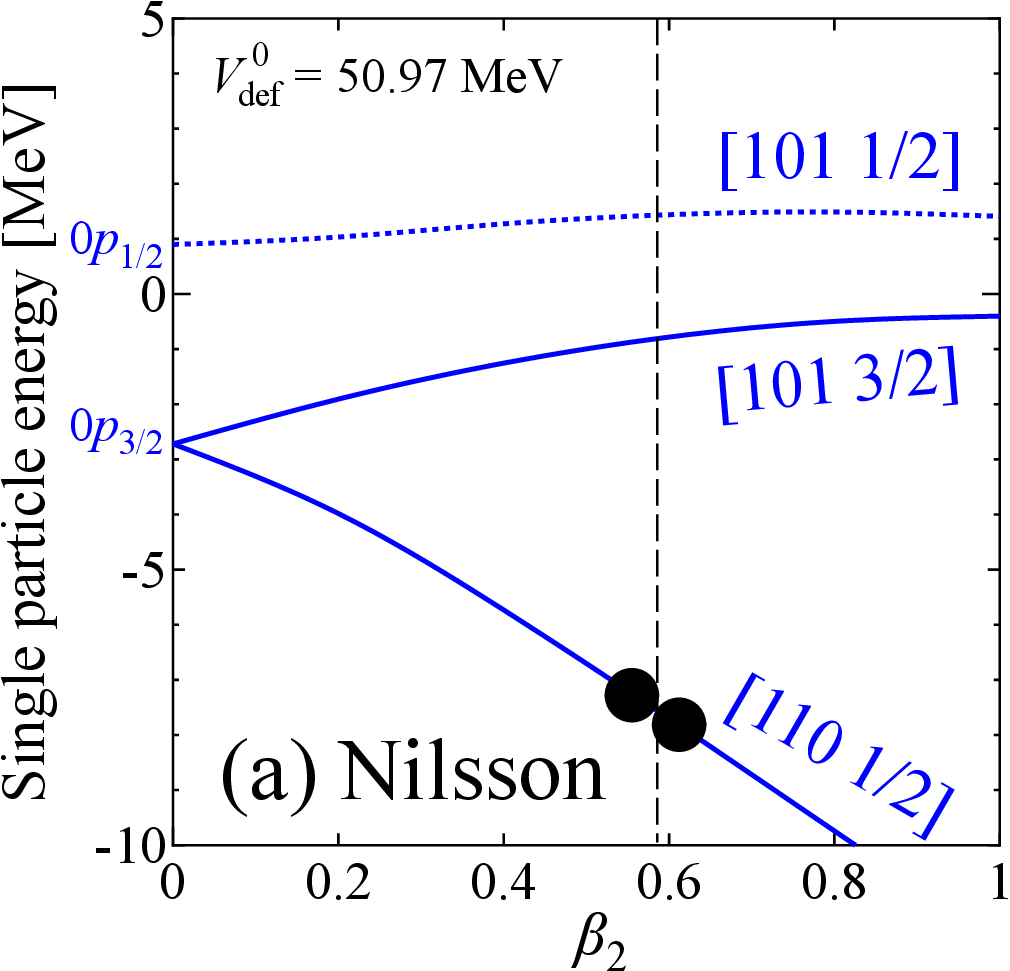}
\end{minipage}
\begin{minipage}{0.33\linewidth}
\includegraphics[width=0.9\linewidth,clip]{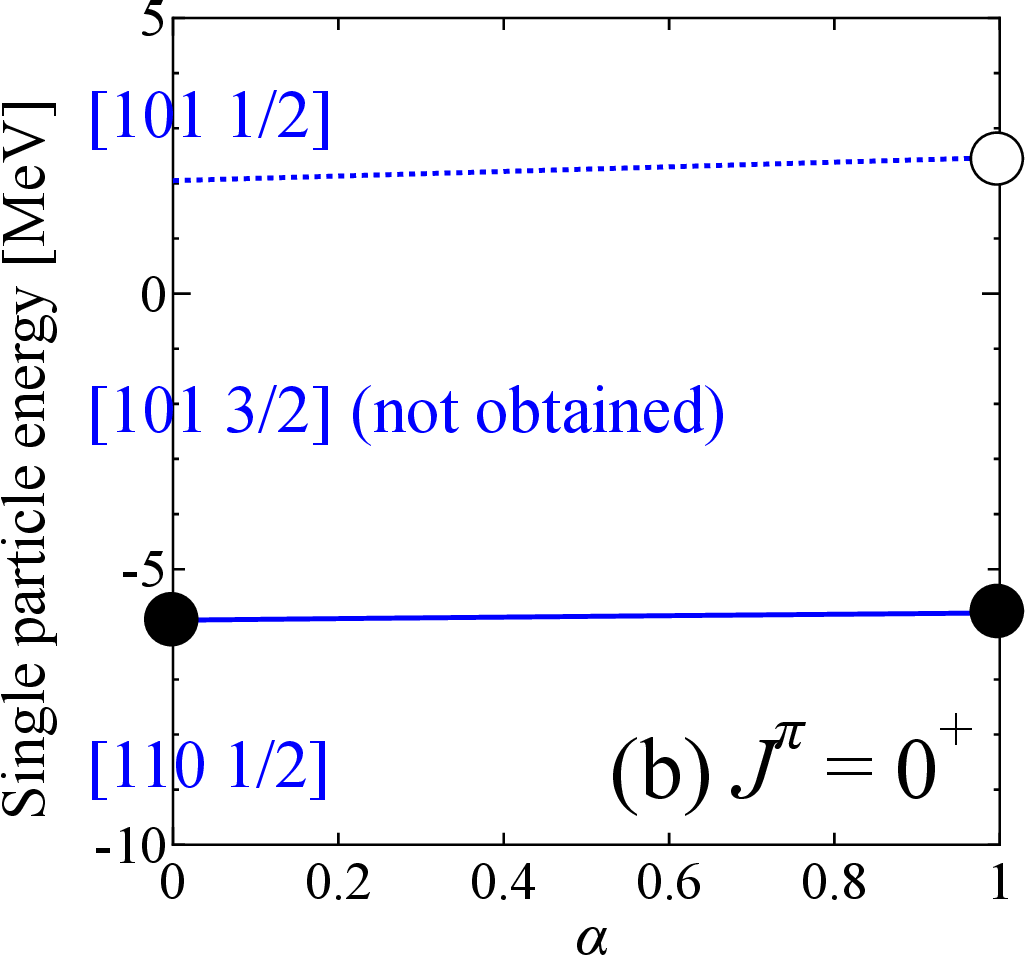}
\end{minipage}
\begin{minipage}{0.33\linewidth}
\includegraphics[width=0.9\linewidth,clip]{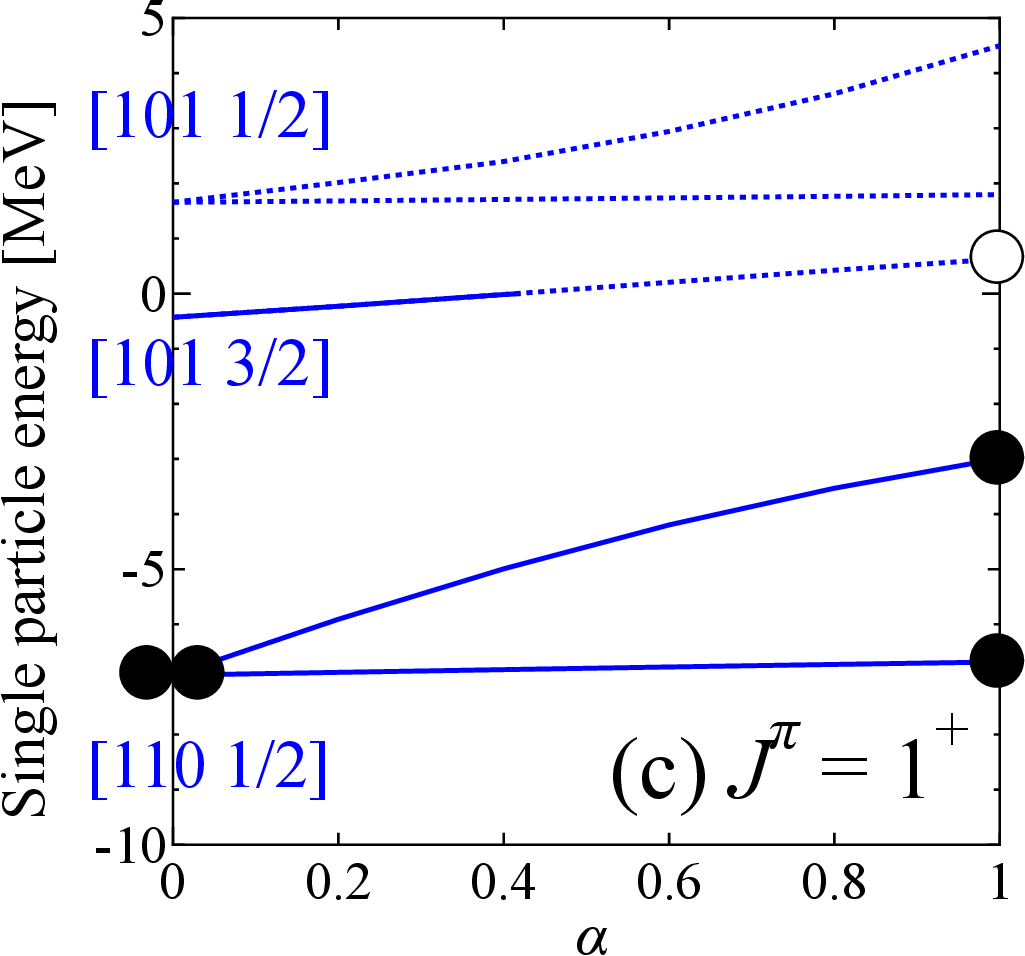}
\end{minipage}

\hfill 
\begin{minipage}{0.33\linewidth}
\end{minipage}
\begin{minipage}{0.33\linewidth}
\includegraphics[width=0.9\linewidth,clip]{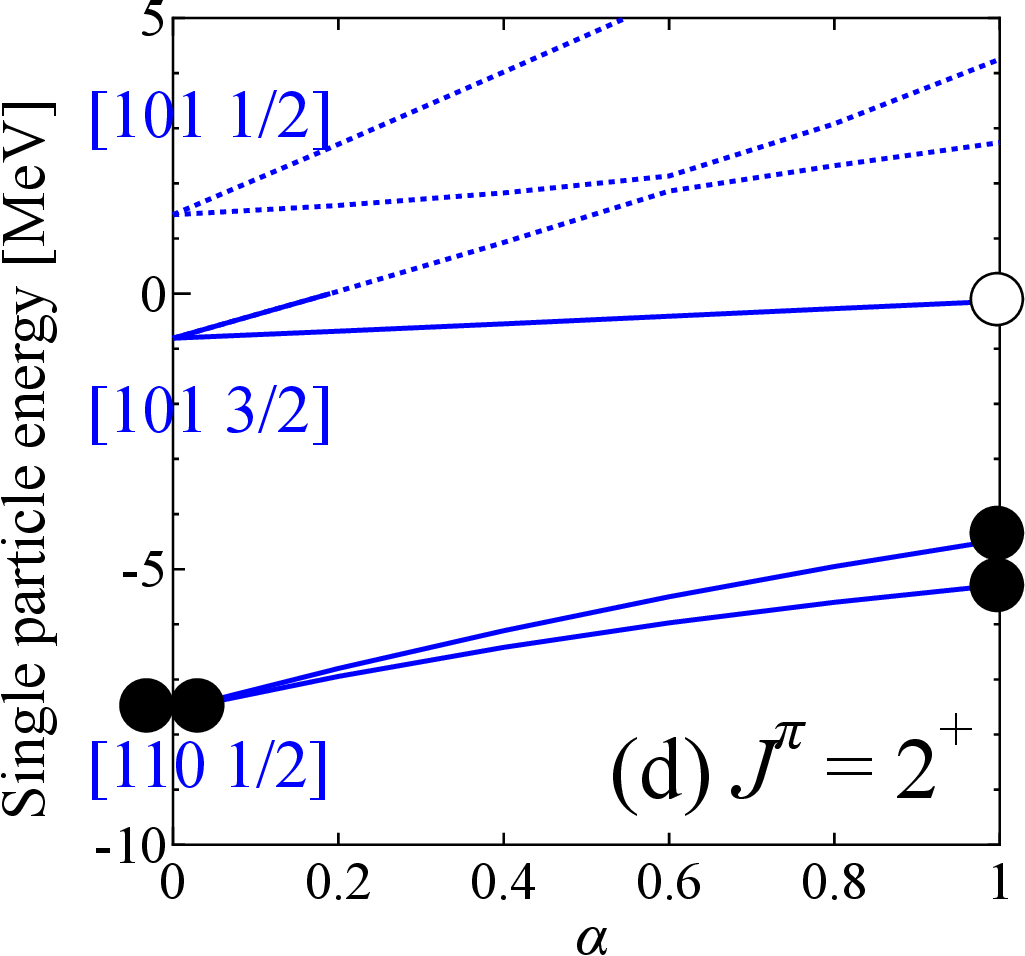}
\end{minipage}
\begin{minipage}{0.33\linewidth}
\includegraphics[width=0.9\linewidth,clip]{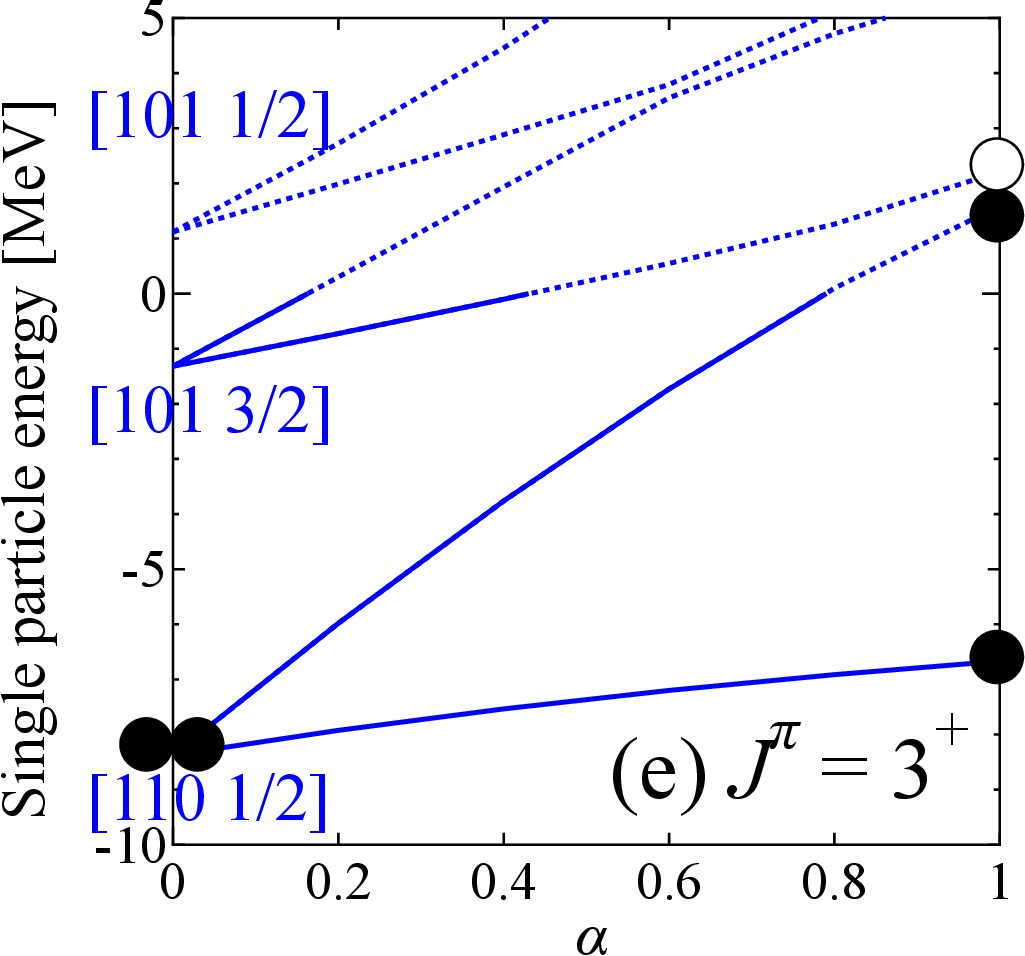}
\end{minipage}
\caption{Single particle energy used to define the forbidden states within the PRM framework.
(a) displays the Nilsson diagram, where the [110 1/2] Nilsson orbital is occupied
by two protons, represented by the filled circles. The value $\beta_2=0.586$ is indicated by the dashed line.
(b)--(e) illustrate the energy splitting as $\alpha$ varies from 0 (Nilsson) to 1 (std-PRM)
for the $J^\pi=0^+$ to $3^+$ states, respectively, with $\beta_2=0.586$.
Note that the potential depths of $V_\mathrm{def}$ are optimized
for each $J^\pi$ as summarized in Table~\ref{tbl:parameters}.
}
\label{fig:nilsson}
\end{figure*}

First, we solve the Schr\"{o}dinger equation using $H_\mathrm{Nil}$ to define the PF states.
Figure~\ref{fig:nilsson}(a) presents the Nilsson diagram obtained within the PRM framework.
The resonant energies indicated by the dotted lines are estimated using the complex scaling method~\cite{Agu71,Aoy06}. In the standard Nilsson model~\cite{Nil55,Rag95}, each orbital is doubly degenerate.
In contrast, in our calculations, the number of Nilsson states depends on $J^\pi$ and the model space $\{\ell j I\}$.
For example, for $J^\pi=0^+$, only one state is obtained for the [110 1/2] Nilsson orbital
because only the $[\mathcal{Y}_{p 3/2}\otimes \phi_{3/2}]_{0^+}$ configuration is allowed with $p_{3/2}$.
Note that the energies corresponding to a specific Nilsson orbital are degenerate for different $J^\pi$. 
This degeneracy allows us to identify which state corresponds to which Nilsson orbital.
Consequently, states up to the [110 1/2] orbital are considered as the PF states in Model III.
For the negative-parity states ($J^\pi=1^-$ and $2^-$), the [000 1/2] Nilsson states
are regarded as the PF states.

On the other hand, in Model I, we further trace how the Nilsson states evolve
by adding $\alpha h_\mathrm{core}$ to $H_\mathrm{Nil}$.
Figures~\ref{fig:nilsson}(b)--\ref{fig:nilsson}(e) illustrate
the energy splitting for $J^\pi=0^+$ to $3^+$, respectively.
This procedure enables us to identify the states originating from the [110 1/2] Nilsson orbital,
which are considered to be PF states. Subsequently, the PF states are
extracted after solving the Schr\"{o}dinger equation with $H_1$.
However, for the $J^\pi=3^+$ state, one of the PF states significantly
increases in energy and approaches the next energy level in the continuum region.
This situation makes it difficult to unambiguously identify the PF state from the solution.
We therefore show the results without excluding this PF state in Model I.

\begin{figure*}[ht]
\centering
\begin{minipage}{0.33\linewidth}
\includegraphics[width=0.9\linewidth,clip]{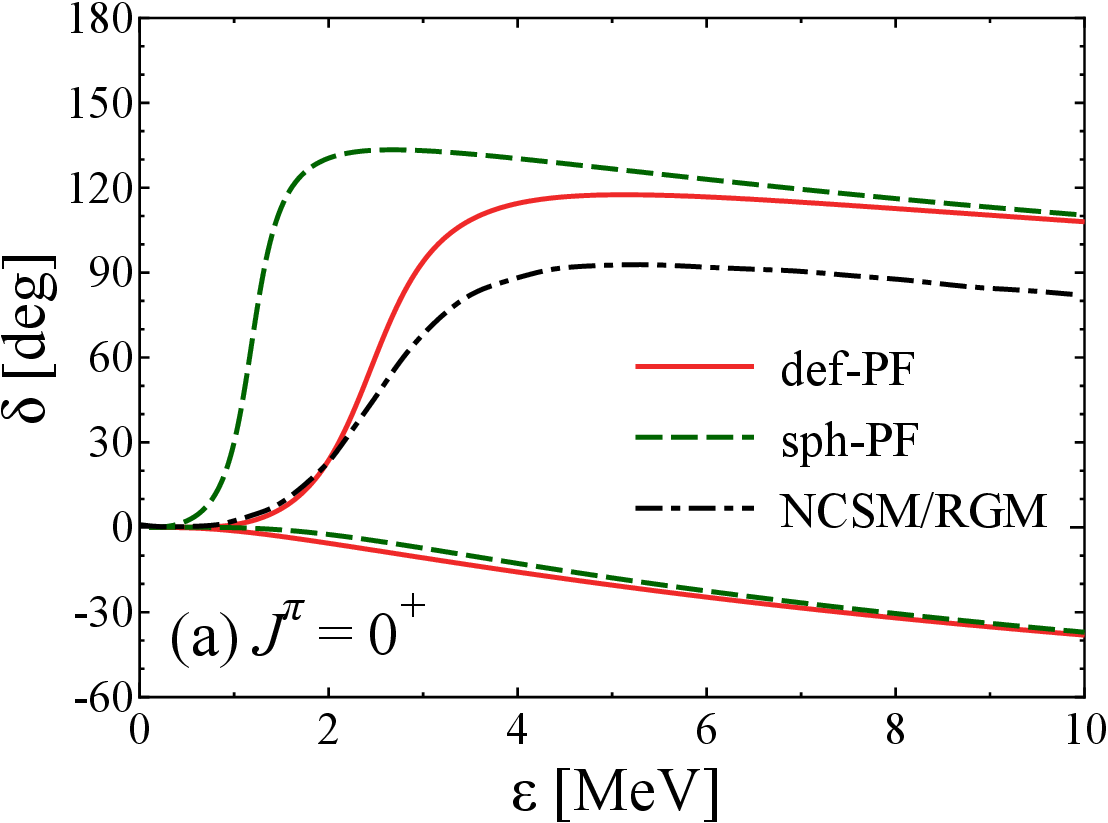}
\end{minipage}
\begin{minipage}{0.33\linewidth}
\includegraphics[width=0.9\linewidth,clip]{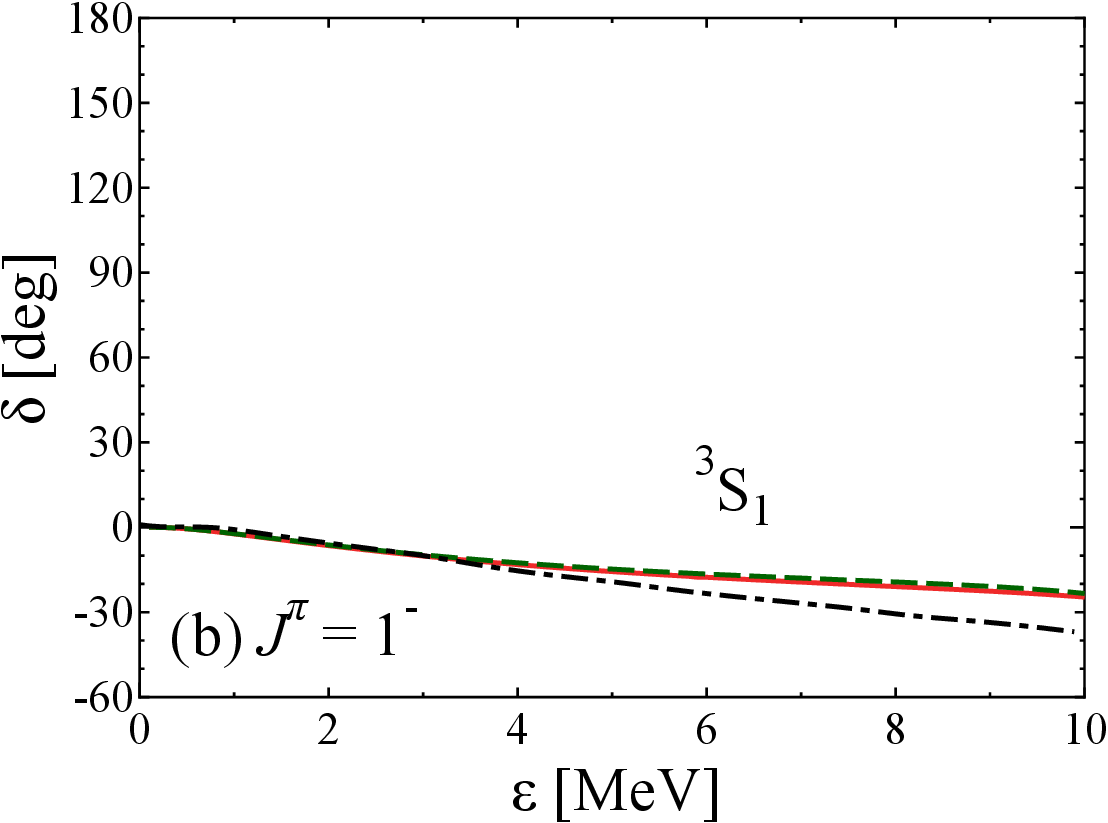}
\end{minipage}
\begin{minipage}{0.33\linewidth}
\includegraphics[width=0.9\linewidth,clip]{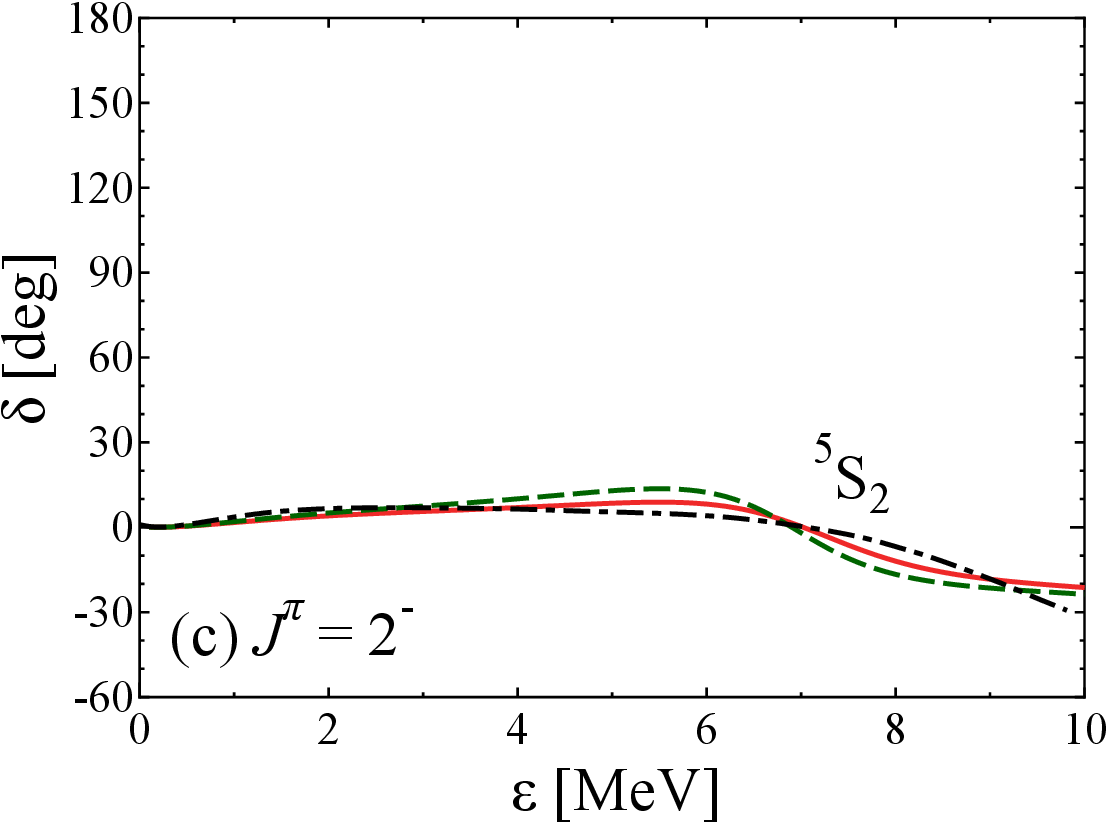}
\end{minipage}
\caption{Eigenphase shift for $J^\pi=0^+$, and diagonal phase shifts for $J^\pi=1^-$ and $2^-$ in the channel-spin representation.
The solid and dashed lines represent Model III (def-PF) and Model II (sph-PF),
while the dot-dashed line corresponds to the ab initio NCSM/RGM calculation~\cite{Kra23}.
It should be noted that Model I (std-PRM) aligns with the solid line for these $J^\pi$.
}
\label{fig:dglps}
\end{figure*}

Next, we determine the potential depth $V_\mathrm{def}^0$ for each $J^\pi$ configuration through the following procedure.
For the $J^\pi=2^+$, $1^+$, and $3^+$ states,
$V_\mathrm{def}^0$ is adjusted to the experimentally well-established bound and resonant energies:
$\varepsilon_\mathrm{gs}(2_1^+)=-0.137$ MeV ($E_x=0$ MeV),
$\varepsilon_\mathrm{res}(1_1^+)=0.63$ MeV ($E_x=0.77$ MeV),
and $\varepsilon_\mathrm{res}(3_1^+)=2.18$ MeV ($E_x=2.32$ MeV).
As for the $3^+$ states in Model I, we assume that the $3^+_1$ state  corresponds to the PF state
and that the $3^+_2$ state is the physical one, which is adjusted to the experimental data.
For the $J^\pi=0^+$, $1^-$, and $2^-$ states,
$V_\mathrm{def}^0$ is tuned so as to reproduce the phase shift calculated by
the ab initio no-core shell model/resonating group method (NCSM/RGM)~\cite{Kra23}.
Figures~\ref{fig:dglps}(a) $J^\pi=0^+$, \ref{fig:dglps}(b) $J^\pi=1^-$, and \ref{fig:dglps}(c) $J^\pi=2^-$
display the resultant phase shifts. The solid lines represent the results of the def-PF model,
which fairly reproduce the NCSM/RGM calculations shown by the dot-dashed lines.
Note that the results of the std-PRM are almost identical.
Similarly, the sph-PF model (dashed line) yields almost the identical phase shifts for the $1^-$ and $2^-$ states.
For the $0^+$ state, the resultant resonant energy $\varepsilon(0_1^+ )\approx1.1$ MeV is somewhat
lower than the NCSM/RGM calculation $\varepsilon(0_1^+ )\approx2.8$ MeV.
However, adjusting this resonance by simply reducing $V_\mathrm{def}^0$ is impractical
as the sph-PF state becomes unbound before reaching the proper position of the $0_1^+$ resonance.
This result supports the use of the Nilsson state as the PF state.
Furthermore, since the $0^+$ state does not significantly impact the elastic
and inelastic cross sections, we have maintained the same potential for the $0^+$ state as used in the other models.
The determined $V_\mathrm{def}^0$ values are summarized in Table~\ref{tbl:parameters}.
It is interesting to note that the obtained parameters remain almost constant
for all $J^\pi$ states, contrasting with the spherical model~\cite{Spa21},
where the potential depths are shallower for the positive-parity states ($V^0\approx40$ MeV)
and deeper for the negative-parity states ($V^0\approx60$ MeV).
This improvement can be understood from the single-particle energy in the Nilsson model.
With deformation, the [101 3/2] Nilsson state increases in energy for the positive-parity
 states (requiring a deeper potential to obtain the same eigenenergy),
while the [220 1/2] Nilsson state decreases for negative-parity states (requiring a shallower potential to achieve the same eigenenergy).

\begin{table}[htbp]
\caption{Potential depth parameters, $V_\mathrm{def}^0$ in MeV, used in the calculations.
}
\label{tbl:parameters}
\begin{center}
\begin{tabular*}{70mm}{@{\extracolsep{\fill}}cccc}
\hline\hline
$J^\pi$
& \begin{tabular}[c]{@{}c@{}}Model I      \\ (std-PRM)\end{tabular}
& \begin{tabular}[c]{@{}c@{}}Model II     \\ (sph-PF) \end{tabular}
& \begin{tabular}[c]{@{}c@{}}Model III    \\ (def-PF) \end{tabular} \\
\hline
 $0^+$  & 48.00  &  48.00  &  48.00  \\
 $1^+$  & 49.83  &  49.23  &  48.70  \\
 $2^+$  & 50.97  &  50.72  &  50.41  \\
 $3^+$  & 52.41  &  50.61  &  49.98  \\
 $1^-$  & 48.00  &  48.00  &  48.00  \\
 $2^-$  & 53.00  &  53.00  &  53.00  \\
\hline\hline
\end{tabular*}
\end{center}
\end{table}

\begin{figure*}[ht]
\centering
\begin{minipage}{0.33\linewidth}
\includegraphics[width=0.9\linewidth,clip]{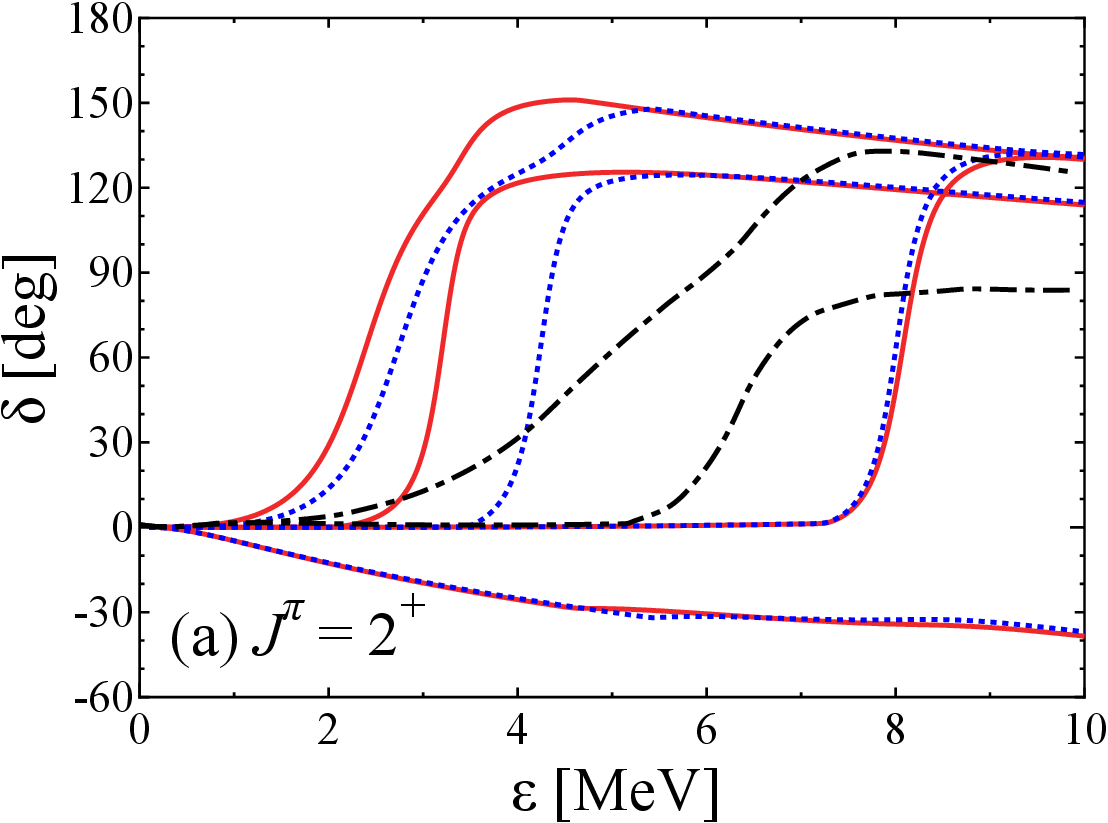}
\end{minipage}
\begin{minipage}{0.33\linewidth}
\includegraphics[width=0.9\linewidth,clip]{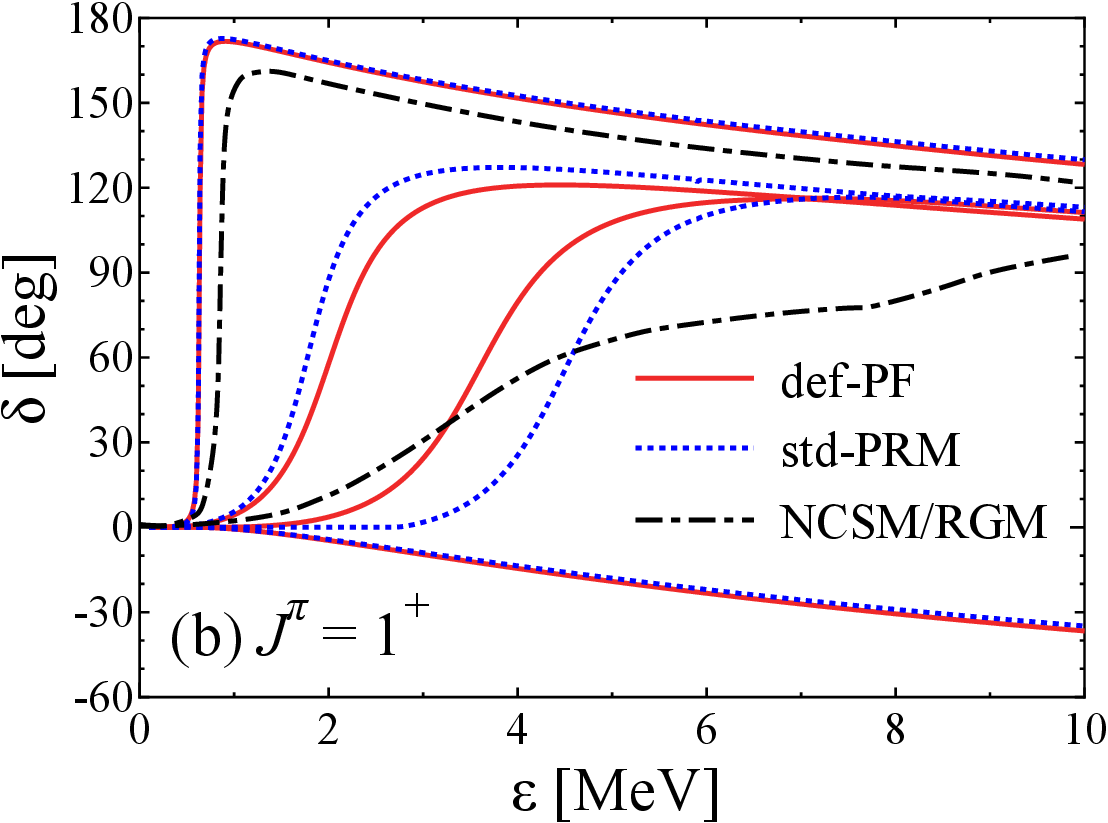}
\end{minipage}
\begin{minipage}{0.33\linewidth}
\includegraphics[width=0.9\linewidth,clip]{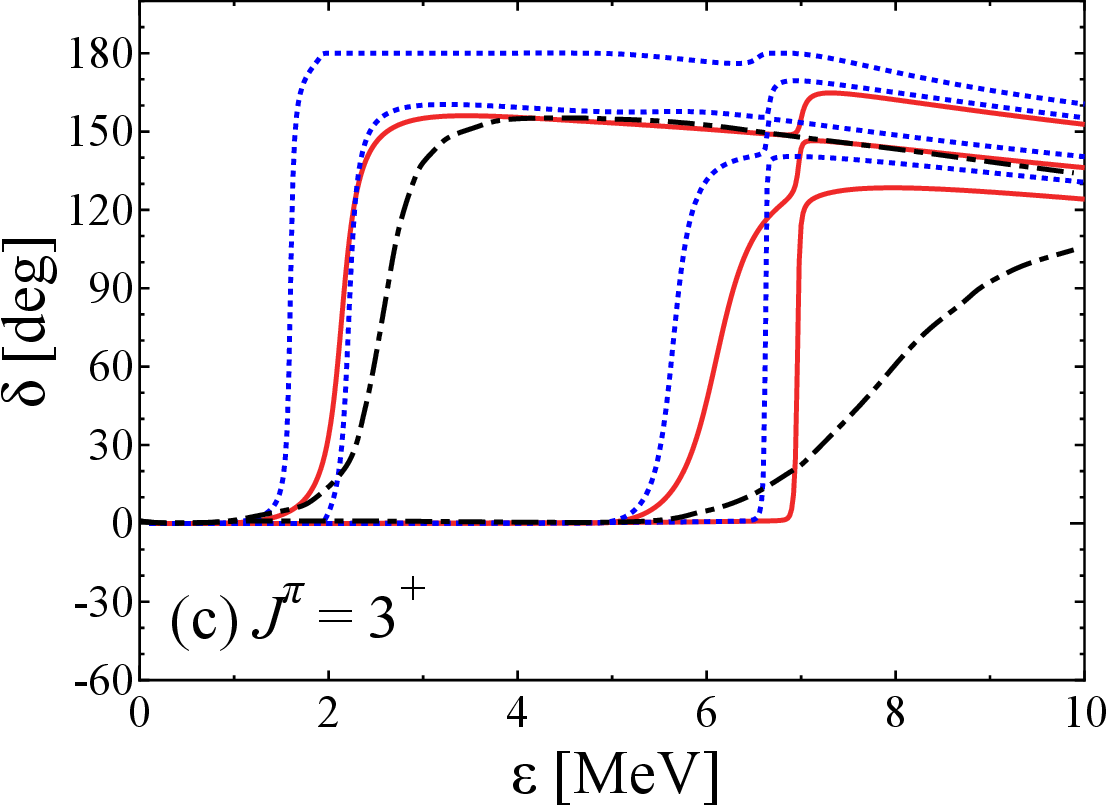}
\end{minipage}
\caption{Eigenphase shifts for $J^\pi=2^+$, $1^+$, and $3^+$.
The solid and dotted line represent Model III (def-PF) and Model I (std-PF),
while the dot-dashed line corresponds to the ab initio NCSM/RGM calculation~\cite{Kra23}.
}
\label{fig:eigenps}
\end{figure*}

Next, we compare the eigenphases calculated with Model I (std-PRM) and Model III (def-PF).
Figure~\ref{fig:eigenps} illustrates the eigenphases for the (a) $J^\pi=2^+$, (b) $J^\pi=1^+$, and (c) $J^\pi=3^+$ states.
The solid and dotted lines represent the def-PF model and the std-PRM, respectively.
Unlike the phase shifts for $J^\pi=0^+$, $1^-$, and $2^-$ depicted in Fig.~\ref{fig:dglps},
$V_\mathrm{PF}$ alters the resonant energies and widths.
Furthermore, for the $3^+$ state in the std-PRM [dotted line in Fig.~\ref{fig:eigenps}(c)],
an additional resonance emerges at around $\varepsilon\approx1.6$ MeV,
which can be regarded as the PF state. 
This resonance cannot be eliminated by changing $V_\mathrm{def}^0$
if the neighboring $3^+$ resonant state is correctly positioned.
For comparison, the NCSM/RGM calculations~\cite{Kra23} are also shown by the dashed lines.
It should be noted that the potential depths are adjusted to the experimental ground
and resonant energies in our calculations (both std-PRM and def-PF),
whereas no phenomenological correction has been applied in the NCSM/RGM.
For the resonances with a narrow width ($J^\pi=1^+$ and $3^+$), the observed behaviors are almost identical.
However, for the resonances with a broader width, the present models do not exhibit
the similar behavior regardless of the inclusion of $V_\mathrm{PF}$.

We analyze the components of the ${}^{8}\mathrm{B}$ wave functions in terms of core and proton states.
Table \ref{tbl:prob-PhysSt} displays the probabilities of $[I^\pi\otimes (\ell j)]$ configurations
in the wave functions for the (a) $J^\pi=2^+$ ground state, (b) $J^\pi=1^+$ resonant state, and 
(c) $J^\pi=3^+$ resonant state using the std-PRM and the def-PF model.
Note that only main components are listed in the tables, with the largest two components highlighted in bold.
For the $2^+$ ground state and the $1^+$ resonant state, the largest two components are the same
although their values slightly differ. However, for the $3^+$ resonant state,
the two dominant components change significantly compared to those of the $2^+$ and $1^+$ states.
This difference may arise from the avoided crossing in the continuum region.
In Fig.~\ref{fig:nilsson} (e), the $3^+$ resonant state of interest, indicated by the open circle,
is significantly approached by the PF state originating from the [110 1/2] Nilsson state.
This proximity leads to a notable exchange of their main components around $\alpha \approx 1$.
Thus, the $3^+$ resonant state is strongly affected by the PF state,
highlighting the importance of adopting alternative approaches to effectively exclude PF states.

Following the analysis of physical states, we further investigate the components
of the ${}^{8}\mathrm{B}$ wave functions for the PF states.
Table \ref{tbl:prob-PF} presents the main configurations $[I^\pi\otimes (\ell j)]$
for the PF states of the (a) $J^\pi=2^+$, (b) $J^\pi=1^+$, and (c) $J^\pi=3^+$
in the Nilsson model and the std-PRM.
The lowest and the second-lowest positive-parity PF states are represented
 as PF1 and PF2, respectively. These states, PF1 and PF2,
correspond to the two leftmost states (Nilsson) and the two rightmost states (std-PRM)
states depicted by the filled circles in Figs.~\ref{fig:nilsson}(c)--\ref{fig:nilsson}(e).
In the Nilsson model calculations, the small value of $\alpha=0.0001$ is used
to effectively resolve the degeneracy of the two [110 1/2] Nilsson states at $\alpha=0$;
the same technique is used to define the sph-PF states (see Sec.~\ref{sec:sph-PF} in detail).
Table \ref{tbl:prob-PF} highlights that incorporating core excitations ($\alpha=1$) generally leads to a
reduction in the high-$I$ components across all the PF states.
Notably, the PF1 state in $J^\pi=1^+$ remains almost unchanged
between the Nilsson model and the std-PRM,
likely due to its minimal high-$I$ components.
Conversely, the significant energy rise for the PF2 state in $J^\pi=3^+$ 
can be attributed to the high proportion (0.688) of the $[9/2^-\otimes p_{3/2}]$
configuration, with $\epsilon_{9/2}=14.6$~MeV, resulting in its emergence in the continuum region.
This situation underscores the necessity
for developing methods to effectively exclude PF states from continuum region.

\begin{table*}[htbp]
\caption{The probability of $[I^\pi\otimes (\ell j)]$
 in the ${}^{8}\mathrm{B}$ wave functions for the (a) $J^\pi=2^+$ ground state,
(b) $J^\pi=1^+$ resonant state, and (c) $J^\pi=3^+$ resonant state
using the std-PRM and the def-PF model.
Only main components are listed in the tables, and the largest two components are shown in bold.
Note that the probabilities for the resonant states are estimated as those of pseudostates using
 the Gaussian expansion method~\cite{Hiy03}.}
\label{tbl:prob-PhysSt}
\begin{center}

(a) $J^\pi=2^+$ ground state\\
\vspace{2mm}
\begin{tabular}{cccccc}
\hline\hline
 Model & $\varepsilon$ (MeV) & $[3/2^-\otimes p_{1/2}]$ & $[3/2^-\otimes p_{3/2}]$ & $[1/2^-\otimes p_{3/2}]$ & $[7/2^-\otimes p_{3/2}]$  \\
\hline
 std-PRM & $-0.137$ & 0.124 & {\bf 0.610} & {\bf 0.141} & 0.113  \\
 def-PF  & $-0.137$ & 0.033 & {\bf 0.756} & {\bf 0.146} & 0.055  \\
\hline\hline
\end{tabular}
\vspace{5mm}

(b) $J^\pi=1^+$ resonant state\\
\vspace{2mm}
\begin{tabular}{cccccc}
\hline\hline
 Model & $\varepsilon$ (MeV) & $[3/2^-\otimes p_{3/2}]$ & $[1/2^-\otimes p_{1/2}]$ & $[1/2^-\otimes p_{3/2}]$ & $[5/2^-\otimes p_{3/2}]$  \\
\hline
 std-PRM & 0.63 & {\bf 0.302} & 0.225 & {\bf 0.339} & 0.094  \\
 def-PF  & 0.63 & {\bf 0.454} & 0.053 & {\bf 0.446} & 0.037  \\
\hline\hline
\end{tabular}
\vspace{5mm}

(c) $J^\pi=3^+$ resonant state\\
\vspace{2mm}
\begin{tabular}{cccccccc}
\hline\hline
 Model & $\varepsilon$ (MeV) & $[3/2^-\otimes p_{3/2}]$ & $[7/2^-\otimes p_{1/2}]$ & $[7/2^-\otimes p_{3/2}]$ & $[5/2^-\otimes p_{1/2}]$ & $[5/2^-\otimes p_{3/2}]$ & $[9/2^-\otimes p_{3/2}]$  \\
\hline
 std-PRM & 2.18 & {\bf 0.244} & 0.078 & 0.175 & 0.111 & {\bf 0.284} & 0.095  \\
 def-PF  & 2.18 & {\bf 0.579} & 0.030 & {\bf 0.364} & 0.000 & 0.012 & 0.000  \\
\hline\hline
\end{tabular}

\end{center}
\end{table*}

\begin{table*}[htbp]
\caption{The probability of $[I^\pi\otimes (\ell j)]$
 in the ${}^{8}\mathrm{B}$ wave functions for the (a) $J^\pi=2^+$,
(b) $J^\pi=1^+$, and (c) $J^\pi=3^+$ PF states (unphysical states)
using the Nilsson model and the std-PRM.
The lowest and the second-lowest positive-parity PF states are represented as PF1 and PF2, respectively.
Only main components are listed in the tables.}
\label{tbl:prob-PF}
\begin{center}

(a) $J^\pi=2^+$ PF states\\
\vspace{2mm}
\begin{tabular}{cccccccc}
\hline\hline
 State & Model & $\varepsilon$ (MeV) & $[3/2^-\otimes p_{1/2}]$ & $[3/2^-\otimes p_{3/2}]$ & $[1/2^-\otimes p_{3/2}]$ & $[7/2^-\otimes p_{3/2}]$ & $[5/2^-\otimes p_{3/2}]$  \\
\hline
 \multirow{2}{*}{PF1} & Nilsson & $-7.56$ & 0.054 & 0.165 & 0.188 & 0.387 & 0.087  \\
                      & std-PRM & $-5.29$ & 0.017 & 0.264 & 0.505 & 0.074 & 0.065  \\
\hline
 \multirow{2}{*}{PF2} & Nilsson & $-7.56$ & 0.101 & 0.000 & 0.226 & 0.322 & 0.279  \\
                      & std-PRM & $-4.47$ & 0.276 & 0.093 & 0.191 & 0.365 & 0.057  \\
\hline\hline
\end{tabular}
\vspace{5mm}

(b) $J^\pi=1^+$ PF states\\
\vspace{2mm}
\begin{tabular}{cccccccc}
\hline\hline
 State &  Model & $\varepsilon$ (MeV) & $[3/2^-\otimes p_{1/2}]$ & $[3/2^-\otimes p_{3/2}]$ & $[1/2^-\otimes p_{1/2}]$ & $[1/2^-\otimes p_{3/2}]$ & $[5/2^-\otimes p_{3/2}]$  \\
\hline
 \multirow{2}{*}{PF1} & Nilsson & $-6.92$ & 0.136 & 0.463 & 0.018 & 0.367 & 0.000  \\
                      & std-PRM & $-6.68$ & 0.141 & 0.491 & 0.018 & 0.338 & 0.000  \\
\hline
 \multirow{2}{*}{PF2} & Nilsson & $-6.92$ & 0.018 & 0.035 & 0.136 & 0.048 & 0.746  \\
                      & std-PRM & $-2.99$ & 0.046 & 0.151 & 0.267 & 0.207 & 0.315  \\
\hline\hline
\end{tabular}
\vspace{5mm}

(c) $J^\pi=3^+$ PF states\\
\vspace{2mm}
\begin{tabular}{ccccccccc}
\hline\hline
 State &  Model & $\varepsilon$ (MeV) & $[3/2^-\otimes p_{3/2}]$ & $[7/2^-\otimes p_{1/2}]$ & $[7/2^-\otimes p_{3/2}]$ & $[5/2^-\otimes p_{1/2}]$ & $[5/2^-\otimes p_{3/2}]$ & $[9/2^-\otimes p_{3/2}]$  \\
\hline
 \multirow{2}{*}{PF1} & Nilsson & $-8.37$  & 0.495 & 0.153 & 0.304 & 0.004 & 0.026 & 0.000  \\
                      & std-PRM & $-6.67$  & 0.752 & 0.086 & 0.136 & 0.001 & 0.008 & 0.000  \\
\hline
 \multirow{2}{*}{PF2} & Nilsson & $-8.37$  & 0.000 & 0.004 & 0.010 & 0.153 & 0.127 & 0.688  \\
                      & std-PRM & $1.63^*$ & 0.182 & 0.001 & 0.392 & 0.147 & 0.153 & 0.111  \\
\hline\hline
\end{tabular}

\end{center}
\noindent $^*$The probabilities of this state are estimated as those of a pseudostate.
\end{table*}

\subsection{Application to the $p+{}^{7}\mathrm{Be}$ reaction}
\begin{figure}[htbp]
\includegraphics[width=0.9\linewidth,clip]{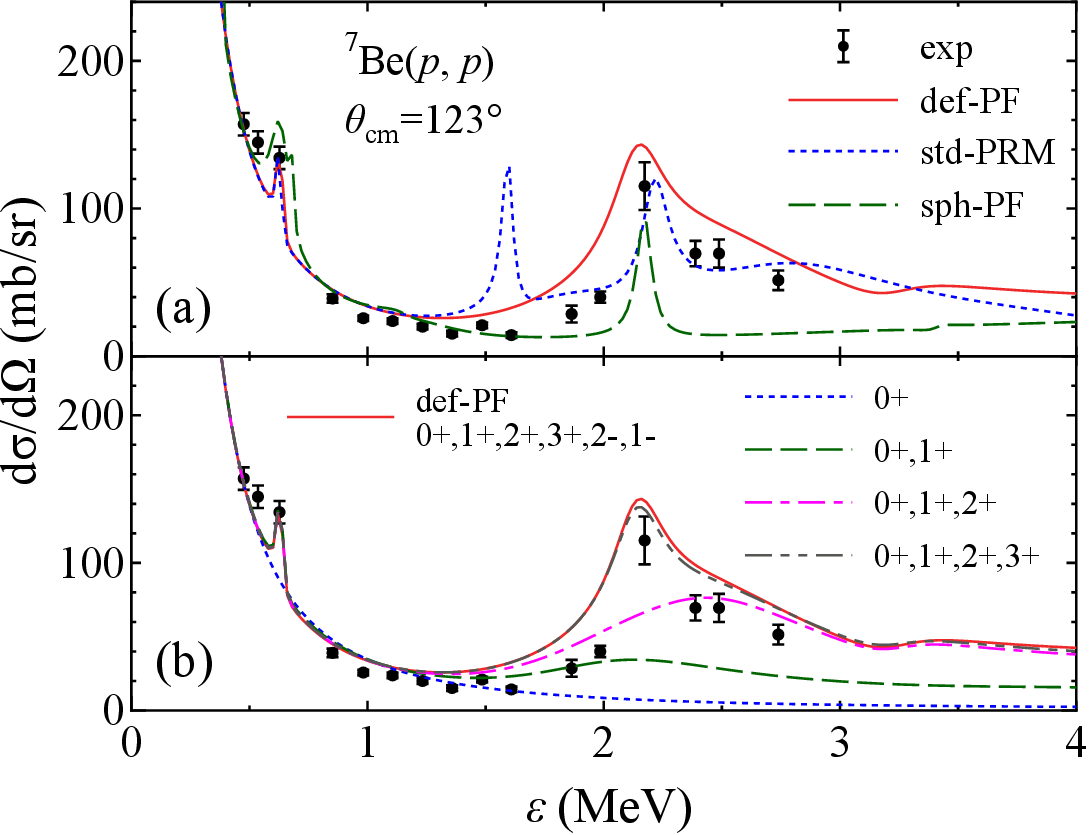}
\caption{Excitation function of $p+{}^7\mathrm{Be}$ elastic scattering cross section at $123^\circ$.
In (a), three models are compared: std-PRM (dotted line), def-PF model (solid line), and sph-PF model (dashed line).
In (b), the $J^\pi$ contributions are illustrated by expanding the model space for the def-PF model (our best result).
The experimental data are taken from Ref.~\cite{Pan19}.
}
\label{fig:el_PRM_def-PF_sph-PF}
\end{figure}

\begin{figure}[htbp]
\includegraphics[width=0.9\linewidth,clip]{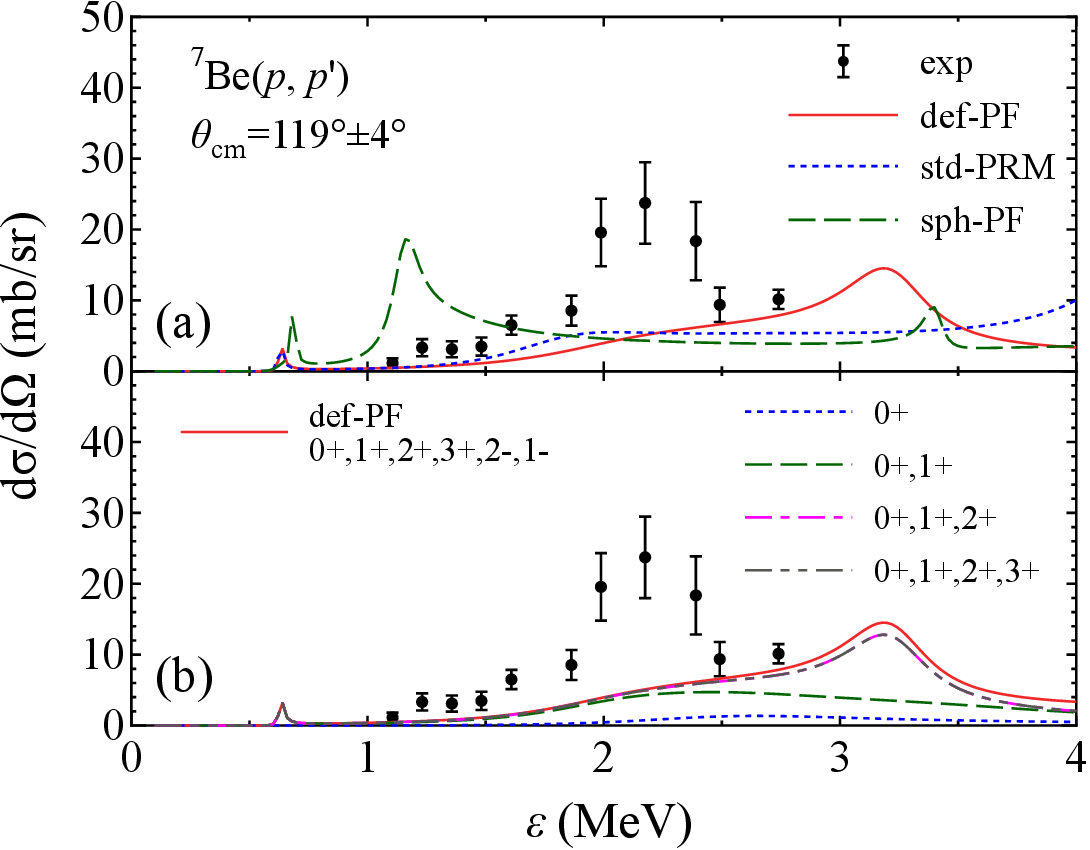}
\caption{Same as Fig~\ref{fig:el_PRM_def-PF_sph-PF},
but for $p+{}^7\mathrm{Be}$ inelastic scattering cross section at $119^\circ$. The experimental data are from Ref.~\cite{Pan19}.
}
\label{fig:inel_PRM_def-PF_sph-PF}
\end{figure}

As a test of the developed models, we have computed excitation functions for the elastic and inelastic cross sections for the $p+{}^7\mathrm{Be}$ reaction, and compared with the existing experimental data \cite{Pan19}.
Figure~\ref{fig:el_PRM_def-PF_sph-PF} shows the excitation function
of the $p+{}^7\mathrm{Be}$ elastic scattering cross section at $123^\circ$.
Figure~\ref{fig:el_PRM_def-PF_sph-PF}(a) compares three models,
while Fig.~\ref{fig:el_PRM_def-PF_sph-PF}(b) illustrates the $J^\pi$ contributions by expanding the model space for our best result, the def-PF model.
In Fig.~\ref{fig:el_PRM_def-PF_sph-PF}(a), the dotted and solid lines represent the results of the PRM and def-PF model, respectively.
Both models exhibit similar behavior in the low-energy region ($\varepsilon\lesssim1.3$ MeV)
and predict  the $1^+_1$ resonance at $\varepsilon=0.63$ MeV,
in agreement with the experimental data~\cite{Pan19}.
Above this energy region, the def-PF model well reproduces the second peak, mainly attributed
to the $3^+$ resonance at $\varepsilon=2.18$ MeV.
On the contrary, as discussed in Fig~\ref{fig:eigenps},
the PRM predicts an additional peak at around $\varepsilon\approx1.6$ MeV, which is inconsistent with the data.
For comparison, the sph-PF model is shown by the dashed line.
The sph-PF model additionally yields two resonances for the $1^+$ states
near the first resonance with $\varepsilon_\mathrm{res}{(1^+_1)}=0.63$ MeV:
$\varepsilon_\mathrm{res}{(1^+_2)}=0.68$ MeV and $\varepsilon_\mathrm{res}{(1^+_3)}=1.13$ MeV.
These two states are primarily composed of the core-excited $I=1/2$ component,
and the corresponding peaks are not observed in the elastic or inelastic (as shown later in Fig.~\ref{fig:inel_PRM_def-PF_sph-PF}) scattering data.
These results underscore the importance of excluding deformed states as PF states
to precisely describe deformed halo nuclei.

Finally, we apply the same model to inelastic scattering.
Figure~\ref{fig:inel_PRM_def-PF_sph-PF} presents the excitation function
for $p+{}^7\mathrm{Be}$ inelastic scattering cross section at $119^\circ$;
the line styles are the same as those used in Fig.~\ref{fig:el_PRM_def-PF_sph-PF}.
In Fig.~\ref{fig:inel_PRM_def-PF_sph-PF}(a), although the peak structure of the def-PF model
 (solid line) more closely approaches the experimental peak
at around $\varepsilon=2.2$~MeV than the PRM (dotted line)
 [see also the phase shift in Fig.~\ref{fig:eigenps}(a)],
 our def-PF model does not adequately reproduce the experimental data.
In Fig.~\ref{fig:inel_PRM_def-PF_sph-PF}(b), the result of the def-PF model is further decomposed into
each $J^\pi$ component.
The decomposition reveals that the inelastic peak is mainly composed of
the $2^+_3$ state at $\varepsilon\approx3.2$ MeV,
which is characterized by the large core-excited component $[\mathcal{Y}_{p 3/2}\otimes \phi_{1/2}]$.
Although the $2^+$ state could be the key to understand the behavior of inelastic scattering,
further refinements of the model are required to accurately understand the complex dynamics.
To preserve the simplicity of the model, we have not attempted such modifications.

\vspace{20mm}

\section{Summary}\label{sec:summary}

In this study, we investigate various methods for excluding PF states in the PRM framework
and evaluate their impact on the resonant-state properties.
Previous studies~\cite{Ura11,Ura12} introduced a simple technique for eliminating PF states
in deformed halo nuclei by removing them after solving the Schr\"{o}dinger equation.
Despite its widespread application, our analysis reveals its limitation for
${}^{8}\mathrm{B}={}^{7}\mathrm{Be}+p$, where a PF state emerges as a resonant state.
To overcome this limitation, we utilize Nilsson states as PF states in the OCM,
allowing us to project them out before solving the Schr\"{o}dinger equation.
This approach successfully reproduces the experimental data on elastic scattering excitation function,
demonstrating its effectiveness.
The same calculation predicts the presence of a low-energy bump in the inelastic scattering
excitation function, although its position is overestimated by about 1 MeV.
This result points out the need for further refinement of this model.
This new approach not only overcomes the challenges encountered in $^8$B calculations
but also broadens the applicability to other nuclei with a deformed core, such as $^{17,19}$C or $^{31}$Ne.
The straightforward integration of this refined projectile wave function with CDCC and other reaction frameworks
promises to open avenues for future research.

\begin{acknowledgments}
We are grateful to P. Punta for providing the benchmark results.
S.W. thanks the faculty and staff at Universidad de Sevilla for their hospitality during
his sabbatical stay, which enabled the completion of this work.
S.W.\ is supported by Japan Society for the Promotion of Science (JSPS) KAKENHI Grant No. JP22K14043.
A.M.M.\ is supported by MCIN/AEI/10.13039/501100011033, grant No.\ PID2020-114687GB-I00. 

\end{acknowledgments}

\bibliography{./ref}

\end{document}